\documentclass[sigconf,screen]{acmart}
\pdfoutput=1
\renewcommand\footnotetextcopyrightpermission[1]{} 
\AtBeginDocument{%
  \providecommand\BibTeX{{%
    \normalfont B\kern-0.5em{\scshape i\kern-0.25em b}\kern-0.8em\TeX}}}
    
\usepackage{amsmath,amssymb,amsfonts}
\usepackage{algorithmic}
\usepackage{graphicx}
\usepackage{textcomp}
\usepackage{xcolor}
\usepackage{multirow}
\usepackage{colortbl}
\usepackage{threeparttable}
\usepackage{tikz}
\usepackage{subcaption}
\usepackage{verbatim}
\usepackage[many]{tcolorbox}
\usepackage[keeplastbox]{flushend} 

\newcommand{\ls}[1]{\textcolor{red}{LS:#1}}
\newcommand{\yx}[1]{\textcolor{blue}{yx:#1}}
\newcommand{\methods}{9\space}

\newcommand{\jjcmt}[1]{\textcolor{red}{#1}}
\renewcommand{\jjcmt}[1]{}
\newcommand{\update}[2]{\textcolor{blue}{#2}}

\newtcolorbox{shadedbox}{
    drop shadow southeast,
    breakable,
    enhanced jigsaw,
    colback=white,
}

\newcounter{finding}

\newcommand{\finding}[1]{\refstepcounter{finding}
    \vspace{2pt}
    \noindent
    \fbox{
        \begin{minipage}{0.455\textwidth}
            \textbf{Finding \arabic{finding}. } \emph{#1}
        \end{minipage}
    }
}

\begin{document}

\title{A Comprehensive Study on Learning-Based PE Malware Family Classification Methods}

\author{Yixuan Ma}
\affiliation{%
  \institution{State Key Laboratory of Communication Content Cognition}
  \city{Tianjin}
  \country{China}}
\affiliation{%
  \institution{College of Intelligence and Computing, Tianjin University}
  \city{Tianjin}
  \country{China}}
\email{yixuan212@tju.edu.cn}

\author{Shuang Liu}
\authornote{Corresponding author}
\affiliation{%
  \institution{College of Intelligence and Computing, Tianjin University}
  \city{Tianjin}
  \country{China}}
\email{shuang.liu@tju.edu.cn}

\author{Jiajun Jiang}
\affiliation{%
  \institution{College of Intelligence and Computing, Tianjin University}
  \city{Tianjin}
  \country{China}}
\email{jiangjiajun@tju.edu.cn}

\author{Guanhong Chen}
\affiliation{%
  \institution{College of Intelligence and Computing, Tianjin University}
  \city{Tianjin}
  \country{China}}
\email{chenguanhong@tju.edu.cn}

\author{Keqiu Li}
\affiliation{%
  \institution{College of Intelligence and Computing, Tianjin University}
  \city{Tianjin}
  \country{China}}
\email{keqiu@tju.edu.cn}

\begin{abstract}
Driven by the high profit, Portable Executable (PE) malware has been consistently evolving in terms of both volume and sophistication. 
PE malware family classification has gained great attention and a large number of approaches have been proposed. With the rapid development of machine learning techniques and the exciting results they achieved on various tasks, machine learning algorithms have also gained popularity in the PE malware family classification task. Three mainstream approaches that use learning based algorithms, as categorized by the input format the methods take, are image-based, binary-based and disassembly-based approaches. 
Although a large number of approaches are published, there is no consistent comparisons on those approaches, especially from the practical industry adoption perspective. Moreover, there is no comparison in the scenario of concept drift, which is a fact for the malware classification task due to the fast evolving nature of malware. 
In this work, we conduct a thorough empirical study on learning-based PE malware classification approaches on 4 different datasets and consistent experiment settings. Based on the experiment results and an interview with our industry partners, we find that (1) there is no individual class of methods that significantly outperforms the others; (2) All classes of methods show performance degradation on concept drift (by an average F1-score of 32.23\%); and (3) the prediction time and high memory consumption hinder existing approaches from being adopted for industry usage. 
\end{abstract}

\vspace{-2mm}



\maketitle

\vspace{-1mm}
\section{Introduction}
\label{sec:introduction}
With the rapid development of information technology, more and more user personal information and property are transferred, stored and even shared on the Internet. Driven by the high profit, underground malware industries have been consistently growing and expanding. It is reported by AVTEST~\cite{avtest} that there are more than 350 thousand new malicious samples registered every day, and the total number of malware has grown roughly 11 times over the past nine years, i.e., from 99 million in 2012 to 1,139 million in 2020. 
Various malware families evolve fast, and tend to attack social and health events to enlarge the damage they can cause. 
According to the 2020 SonicWall Cyber Threat Report~\cite{sonicwall}, over the past half year, 
attacks related to COVID-19 were increased. The biochemical systems at an Oxford university research lab currently studying the Covid-19 pandemic has been breached~\cite{covid-19}.

It is well recognized that malware family evolves fast and the concept drift, which is the change in the statistical properties of an object in unforeseen ways, has become a rather challenging issue in the domain~\cite{Usenix17Transcent}. Malware family classification is of great importance to understand the malware family evolving trend and to better detect malware~\cite{importance}. This is particularly significant for portable executable (PE) malware, as Windows is one of the most widely used operating systems. According to  MalwareBazaar~\cite{malbazaar-stastic}, a website which offers free malware sample upload and download services, and some existing research  findings~\cite{Gibert2019using, vasan2020imcfn}, PE files account for around 50$\%$ of the malware uploaded by user.  A large number of research approaches have been proposed to solve the PE malware classification problem, among which learning-based methods~\cite{Nataraj2011malware, Gibert2019using,vasan2020imcfn} have recently become popular benefiting from the success of machine learning, especially deep learning techniques.

Learning-based PE malware family classification techniques can mainly be categorized into three classes based on the malware file  format adopted. The first class of methods convert PE malware files into images, and then apply image classification techniques to solve the problem~\cite{Nataraj2011malware, LBP2016novel, PCA2016performance, Gibert2019using, vasan2020imcfn}. The second class of methods directly take the PE malware file as a sequence of binary code and usually sequential models from the domain of Natural Language Processing (NLP) are adopted to solve the problem~\cite{jain2011byte, raff2017malware, raff2017learning, qiao2020malware, santos2013opcode, awad2018modeling}. More recently, a new type of methods, which de-compile the PE malware file into assembly code, and then adopt graph structure analysis on the control flow graph (CFG) of the assembly code~\cite{kinable2011malware, kong2013discriminant, hassen2017scalable, yan2019classifying}. 
These three categories of methods are independently evaluated with different datasets, and they all show good performance on the evaluated dataset. The most recent approaches~\cite{vasan2020imcfn} are reported to achieve more than 98\% F1-score on the BIG-15~\cite{big20152018microsoft} dataset. 

There are some survey papers that summarize existing methods of malware family classification~\cite{raff2020survey, gibert2020rise, ucci2019survey, souri2018state} and Iadarola et al.~\cite{iadarola2020image} conduct an evaluation of image-based texture feature extraction methods. 
However, there is still no existing study that systematically evaluates all three classes of learning-based methods on the same evaluation settings with the same datasets. Moreover, 
%
given that those PE malware family classification approaches report excellent performance on existing public of manually crafted  datasets, it is not clear (1) which class of methods are more accurate and efficient; (2) how do different kinds of methods perform in circumstances of concept drift, which is a common and realistic problem for the fast evolving malware family; and more importantly, (3) what is the status of industry adoption of those methods and the future research directions to support industry requirements.


In this work, we seek to answer the questions with a well designed empirical study. We first briefly review existing approaches and select \methods most representative approaches from the three categories: 4 image-based approaches (VGG-16~\cite{vasan2020imcfn}, ResNet-50~\cite{vasan2020imcfn, singh2019malware, rezende2017malicious, bhodia2019transfer}, Inception-V3~\cite{vasan2020imcfn},  IMCFN~\cite{vasan2020imcfn}), 2 binary-based approaches (CBOW+MLP~\cite{qiao2020malware}, MalConv~\cite{raff2017malware, krvcal2018deep}) and 3 disassembly-based approaches (MAGIC~\cite{yan2019classifying}, Word2Vec+KNN~\cite{awad2018modeling, chandak2021comparison}, MCSC~\cite{ni2018malware, kwon2020malware}).

Our results suggest that (1) no individual class of methods  outperforms the others and the binary-based method CBOW+MLP~\cite{qiao2020malware} achieves the best performance across different datasets; (2) All classes of methods show performance degradation, by an average of 32.23\%, in the circumstance of concept drift and CBOW+MLP~\cite{qiao2020malware} shows the sharpest performance degradation (by 69.62\%); (3) Industry mainly adopts sandbox and pattern database to detect malware families currently, and the long prediction time, fragility to malware evolution and the high resource consumption hinders the current learning-based methods from being applied to industry practices. 

The main contributions of this work are as follows: 
\begin{itemize}
    \item We conducted the first large-scale and systematic empirical study on learning-based PE malware family classification methods.
    \item We create two new PE malware family classification datasets, one for the normal classification purpose and one for the concept drift purpose, and we will make them public.
    \item We are the first to conduct evaluations on the concept drift situation with a large number of representative methods.
    \item We provide practical suggestions and future research directions to promote applicable research in the area of malware family classification. 
\end{itemize}

All experimental data and results are publicly available to encourage future research in the community: \textbf{\url{https://github.com/MHunt-er/Benchmarking-Malware-Family-Classification}}.
\section{Related Work}
\label{sec:relatedwork}

\begin{table*}[t]
\setlength{\abovecaptionskip}{0.15cm}
\centering
\footnotesize
\begin{threeparttable}
\caption{The Malware Family Classification Methods Studied}
\label{tab:methods}
\small
\begin{tabular}{lllll}
\toprule

\hline
\textbf{Category}                           & \textbf{Model}                        & \textbf{Related Work}  & \textbf{Input Format for Model} & \textbf{PE Dataset for Evaluation}               \\ \hline
                                            & VGG-16                                & Vasan et al.~\cite{vasan2020imcfn}   & color image                     & Malimg                            \\
                                            & ResNet-50                             & Vasan et al.~\cite{vasan2020imcfn}   & color image                     & Malimg                            \\
                                            &                                       & Singh et al.~\cite{singh2019malware}   & color image                     & Self-constructed dataset$^*$      \\
                                            &                                       & Rezende et al.~\cite{rezende2017malicious} & color image                     & Malimg                            \\
                                            &                                       & Bhodia et al.~\cite{bhodia2019transfer}  & grayscale image                 & Malicia, Malimg                   \\
                                            & Inception-V3                          & Vasan et al.~\cite{vasan2020imcfn}   & color image                     & Malimg                            \\
\multirow{-7}{*}{\textbf{Image-based}}            & IMCFN                                 & Vasan et al.~\cite{vasan2020imcfn}   & grayscale image, color image    & Malimg                            \\ \hline
                                            & {\color[HTML]{000000} CBOW+MLP} & Qiao et al.~\cite{qiao2020malware}    & byte embedding ascending matrix & BIG-15                            \\
                                            & MalConv Related                       & Raff et al.~\cite{raff2017malware}    & raw byte values                 & Closed dataset$^\star$                    \\
\multirow{-3}{*}{\textbf{Binary-based}}      &                                       & Krcál et al.~\cite{krvcal2018deep}   & raw byte values                 & Closed dataset$^\star$                    \\ \hline
                                            & MAGIC                                 & Yan et al.~\cite{yan2019classifying}     & attribute control flow graph    & BIG-15, YANCFG                    \\
                                            & Word2Vec+KNN                       & Awad et al.~\cite{awad2018modeling}    & opcode sequences                & BIG-15                            \\
                                            &                                       & Chandak et al.~\cite{chandak2021comparison} & 20 most frequent opcodes        & Malicia, Self-constructed dataset$^*$ \\
                                            & MCSC                         & Ni et al.~\cite{ni2018malware}      & opcode sequences Simhash image  & BIG-15                            \\
\multirow{-5}{*}{\textbf{Disassembly-based}} &                                       & Kwon et al.~\cite{kwon2020malware}    & opcode sequences Simhash image  & BIG-15                            \\ \hline

\bottomrule
\end{tabular}
\small
\begin{tablenotes}
  \item[$*$] The malware samples of self-constructed dataset comes from various open source data release websites.
  \item[$\star$] The dataset was provided by an anti-virus industry partner with whom the authors work, which cannot be accessed publicly.
\end{tablenotes}
\vspace{-4mm}
\end{threeparttable}
\end{table*}


\subsection{Image-Based Techniques}
Image-based Techniques first transform the malware binary files into either grayscale or color images, and then adopt image classification models for malware family classification. 

Nataraj et al.~\cite{Nataraj2011malware} propose to transform malware binary files into grayscale images, and employ K-Nearest Neighbor (KNN) model for malware classification via leveraging the texture features of images. This is also the first work that visualizes malware files into images for classification purposes.
Since then, many approaches have been proposed to further improve it. For example, Ahmadi et al.~\cite{LBP2016novel} adopt boosting tree classifiers to better capture multiple novel features, such as Haralick and Local Binary Patterns (LBP).
Narayanan et al.~\cite{PCA2016performance} study the performance of various machine learning models with texture features extracted via Principal Component Analysis.
%
However, those methods are typically inefficient due to the high overhead of extracting complex texture features.

With the recent advances of deep learning models in image classification tasks, they also attract attention in the security community, and have been employed for malware family classification. For example, Convolutional Neural Networks (CNN) have been widely adopted by many approaches~\cite{Gibert2019using, vasan2020imcfn, vasan2020imcec, extrmeML2020convolutional, kalash2018malware}, 
and have shown better performance compared to traditional machine learning approaches~\cite{Gibert2019using}.
%
Vasan et al.~\cite{vasan2020imcfn} propose to adopt transfer learning, which fine-tune a CNN network pre-trained with the ImageNet dataset~\cite{imagenet2009imagenet}, with the images of malware files to better conform the malware family classification task.
The empirical results show that their architecture stands out among several deep learning models. 

Image-based approaches require no domain-specific knowledge, and many existing well-developed image classification models can be used directly.
However, they also have drawbacks. For example, transforming malware into images introduces new hyper-parameters (e.g., image width) and imposes non-existing spatial correlations between pixels in different rows which might be  false~\cite{gibert2020rise}. 

\subsection{Binary-Based Techniques}
Binary-based approaches take the binary code of malware as input, which is considered as sequential information. Therefore, existing sequential models, especially from the Natural Language Processing (NLP) domain, are usually adopted for classification. 

Jain et al.~\cite{jain2011byte} propose to extract n–grams from the raw byte sequences as features, and leverage several basic machine learning models like Na\"ive Bayes~\cite{naivebayes2001empirical} and AdaBoost~\cite{adaboost1997decision} for malware detection. However, with the increase of n, the computation cost increases exponentially. Raff et al.~\cite{raff2017malware} propose the first end-to-end shallow CNN model 
that directly takes the entire binary file as input for malware classification. 
As a result, it will requires large memory for large malware files, and thus has limited processing capability. 
To solve the problem, the approach that focuses only on the PE-header of binaries~\cite{raff2017learning} is proposed. It selects 328 bytes from the PE-header as inputs, and thus is not affected by the size of malware files. It is also shown to perform better than the compared approach that depends on domain knowledge features extracted by
PortEX library~\cite{portex2014robust}. 
Qiao et al.~\cite{qiao2020malware} treat each malware binary file as a corpus of 256 words (0x00-0xFF), adopt the Word2Vec model~\cite{word2vec2013efficient} to obtain the word embeddings, then represent the malware as a word embedding matrix in byte ascending order, and finally classify malware using Multi-Layer Perception (MLP).

Binary-based approaches do not require domain knowledge and  consider the contextual information in malware binaries. However, representing a malware sample as a sequence of bytes may present some challenges compared to other category of methods. 
First, by treating each byte as a unit in a byte sequence, the size of the malware byte sequences may reach several million time steps~\cite{raff2017malware}, which is rather resource consuming. 
Second, adjacent bytes are expected be correlated spatially, which may not always hold due to jumps and function calls,
and thus there might be discontinuities in the information within the binary files.

\vspace{-1mm}
\subsection{Disassembly-Based Techniques}
Disassembly-based approaches first disassemble binary files into assembly code, and perform malware classification based on features such as Function Call Graph (FCG) and Control Flow Graph (CFG), that are extracted from assembly code.



Kinable et al.~\cite{kinable2011malware} propose to calculate the similarity score between two FCGs through existing graph matching techniques and use it as the distance metric for malware clustering. 
Kong et al.~\cite{kong2013discriminant} present a generic framework that can first abstract malware into Attribute Function Call Graphs (AFCGs) and then further learn discriminant malware distance metrics to evaluate the similarity between two AFCGs. 
The above approaches are computation-intensive while calculating the similarity between graphs, which will bring huge performance overhead and cannot generalize well. 
Recently, Hassen et al.~\cite{hassen2017scalable} propose to cluster similar functions of FCGs by using Minhash~\cite{minhash1997resemblance}, and then represent the graphs as vectors for classification via leveraging deep learning models.
Similarly, Yan et al.~\cite{yan2019classifying} employ Deep Graph Convolution Neural Network (DGCNN)~\cite{dgcnn2018end} to aggregate the attributes on the AFCGs extracted from disassembly files, which is the first attempt for DGCNN on malware classification tasks.

There are also some approaches which extract features directly from assembly code. Specifically, the opcode sequence is usually adopted. 
Santos et al.~\cite{santos2013opcode} propose a feature representation method based on the frequency of the appearance of opcode sequence.
Awad et al.~\cite{awad2018modeling} treat opcode sequences of each disassembly file as a document and apply the Word2Vec model~\cite{word2vec2013efficient} to generate a computation representation of the document. Finally, they use the Word Mover’s Distance (WMD)~\cite{wmd2015word} metric and K-Nearest Neighbour (KNN) to classify these documents.
SimHash~\cite{ni2018malware} and MinHash~\cite{sun2018deep} adopt Hash projections to convert opcode sequences into vectors, which are then visualized into images for classification. 

The disassembly-based techniques can better capture the code structure features compared to other methods, but they usually require domain knowledge, such as assembly language and its corresponding analysis methods.  


\vspace{-1mm}
\section{Design of Empirical Study}
\label{sec:method}



\subsection{Research Questions}
\label{subsec:rq}

\noindent\textbf{RQ1: How do different PE malware family classification methods perform on different datasets?} 
Although a large number of learning-based approaches are proposed to solve the malware family classification task, they are only evaluated independently with some specific dataset, e.g., public dataset BIG-15~\cite{big20152018microsoft}, some manually crafted dataset~\cite{singh2019malware} or dataset provided by their industry partners which are not public available~\cite{raff2017malware}. There is a lack of systematic study to evaluate the performance of different approaches consistently on the same experiment settings with multiple datasets. 


\noindent\textbf{RQ2: How is the classification performance of various models affected by malware concept drift?} 
Concept drift~\cite{Usenix17Transcent}, which is the change in the statistical properties of an object in 
unforeseen ways, is a realistic and critical issue in the PE malware family classification task. It is thus important to evaluate the performance of different approaches in the  application scenario of concept drift. 

\noindent\textbf{RQ3: What factors hinder the current learning-based PE malware classification approaches from being deployed in industry and the corresponding improvement directions?}
With the gaps identified by the previous research questions, our ultimate goal is to provide suggestions on how to make the current learning-based PE malware classification approaches applicable in real industry scenarios. 

\subsection{Studied Methods }
\label{subsec:techniques}
In order to systematically study the performance of different techniques, we select \methods state-of-the-art  learning-based PE malware family classification methods, which cover image-based, binary-based and disassembly-based techniques, for our empirical study.\update{ Since learning-based methods adopt different machine learning models, which in turn could be adopted by various different methods. We use the corresponding machine learning models, which achieve state-of-the-art performance, as the target of our study to simplify the mapping work. }{}
Table~\ref{tab:methods} lists the details of all methods adopted.\update{including the category of the model, the related work adopting the model, the input formats required by the model and the datasets used for evaluation in the corresponding study. }{}



\subsubsection{Image-Based}



\vspace{1mm}
\noindent\textbf{VGG-16:} 
VGGNet~\cite{vgg2014very} was proposed to reduce the number of parameters in the convolution layers (with smaller convolution kernels) and improve on training time.
VGG-16 is the most popular network architecture in the VGGNet family. The standard VGG-16 contains 13 Convolution layers (CONV), 5 max Pooling layers (Pool) and 3 Fully connected layers ($4096\times4096\times N$, where N is decided by the number of predicted classes). All hidden layers use the Rectified Linear Unit (ReLU). For the output layer, softmax is used. The advantages of VGG-16 are its simple structure and its high fitting capability due to the large number of parameters.

\noindent\textbf{ResNet-50:} ResNet~\cite{resnet2016deep} is proposed to solve the problems of information loss, vanishing gradient and  gradient explosion which commonly exist in the information transmission of traditional CNN networks when the network layers are deep. 
It protects the integrity of information by taking the input directly to the output in a structure called \textit{skip connection (shortcut)}.
Resnet-50 is a typical network in ResNet family. Because of the design of \textit{shortcut}, ResNet-50 can train a deeper network which contains 50 hidden layers. ResNet-50 finally uses the global average Pool, and then uses softmax for the final prediction layer. ResNet-50 is more efficient than VGG-16 because it has far fewer parameters. 

\noindent\textbf{Inception-V3:} InceptionNet~\cite{inception-v1, inception-v2, inception-v32016rethinking, inception-v4} aims to figure out how to use a dense component to approximate or replace an optimal locally sparse convolution structure.
Four versions of InceptionNet have been developed, in which Inception-V3~\cite{inception-v32016rethinking} is the most representative one. It contains both symmetric and asymmetric building blocks, including CONV, average Pool, max Pool, Concatenation, dropout and FCs. Batch-normalization is widely used throughout the model. Like ResNet-50, Inception-V3 also ends up using a global average Pool, and then uses softmax for the final prediction layer. Inception-V3 was a first runner up in the ImageNet Large Visual Recognition Challenge (LVRC) where it outperformed VGGNet on error rate~\cite{lvrc}. It is more effective compared to VGG-16 and it has a even smaller number of parameters than ResNet-50. 

\noindent\textbf{IMCFN:} 
IMCFN stands for Image-based Malware Classification using Fine-tuned Convolutional Neural Network, which is customized to the task of malware family classification. It is a variant of VGG-16 that reduces the neurons in the first two FCs from 4,096 to 2,048, and adds a dropout layer to reduce the effect of overfitting. 


\begin{table}[t]
\setlength{\abovecaptionskip}{0.15cm}
\centering
\footnotesize
\begin{threeparttable}
\setlength\tabcolsep{20pt}
\caption{Details of MalwareBazaar dataset}
\label{tab:new-dataset}
\begin{tabular}{cc}
\toprule
\hline
Family Name & Number of Samples \\ \hline
Gozi        & 767               \\
GuLoader    & 589               \\
Heodo       & 214               \\
IcedID      & 578               \\
njrat       & 942               \\
Trickbot    & 881               \\ \hline
Total       & 3,971              \\ \hline
\bottomrule
\end{tabular}
\vspace{-2mm}
\end{threeparttable}
\end{table}

\begin{table}[t]
\setlength{\abovecaptionskip}{0.15cm}
\centering
\footnotesize
\begin{threeparttable}
\caption{Details of MalwareDrift dataset}
\label{tab:drift-dataset}
\begin{tabular}{ccc}
\toprule
\hline
Family Name & Samples of Pre-drift        & Samples of Post-drift        \\ \hline
Bifrose                      & 171              & 107                \\
Ceeinject                    & 90               & 458                \\
Obfuscator                   & 143              & 61                 \\
Vbinject                     & 379              & 653                \\
Vobfus                       & 64               & 218                \\
Winwebsec                    & 218              & 269                \\
Zegost                       & 180              & 114                \\ \hline
Total & 1,245 & 1,880
\\ \hline
\bottomrule
 \end{tabular}
\end{threeparttable}
\vspace{-5mm}
\end{table}

\vspace{-3mm}
\subsubsection{Binary-Based}


\vspace{1mm}
\noindent\textbf{Word2Vec+MLP: }
This approach combines Word2Vec and the Muli-Layer Perception (MLP) model for malware family classification~\cite{qiao2020malware}. 
The key idea of this method is that the relationship of bytes in samples from the same family is similar, and is distinctly different from those samples of different families. Therefore, the vector matrix of raw bytes is a valid feature for malware classification. 
The raw binary is first pre-processed, removing 5 or more consecutive 0x00 or 0xCC (meaningless bytes). Each file is then taken as a corpus, which is considered to be composed of 256 words from 0x00 to 0xFF. The Continuous Bag-of-Word Model (CBOW)  in  Word2Vec~\cite{word2vec2013efficient} is used to obtain embedding vectors of 256 bytes in the file, and each file is represented as a byte vector ascending matrix. MLP takes these matrices as inputs and outputs the corresponding family categories.
MLP consists of 3 FCs ($512\times512\times N$, where N is decided by the number of predicted classes), and for the first two FCs, a dropout layer is added. MLP is the most intuitive and simplest deep neural network. Similar to VGG-16, although its structure is relatively simple, it has many parameters and can fit the training data well. 

\vspace{1mm}
\noindent\textbf{MalConv:} 
This is the first end-to-end malware detection model that allows the entire malware to be taken as input. 
MalConv~\cite{raff2017malware} first uses an embedding layer to map the raw bytes to a fixed 8 dimensional vector. 
In this way, it can captures high level location invariance in raw binaries by considering both local and global contexts. Then, MalConv use a shallow CNN with a large filter width of 500 bytes combined with an aggressive stride of 500. This allowed the model to better balance computational workload in a data-parallel manner using PyTorch~\cite{pytorch} and thus can relieve the problem of the GPU memory consumption in the first convolution layer. 
As a shallow CNN architecture,  MalConv conquers one of the primary practical limitations that reading the whole malware bytes is memory consuming, it also captures global  location invariance in raw binaries. It allows the embedding layer to be trained jointly with the convolution layer for better feature extraction. 

\vspace{-2mm}
\subsubsection{Disassembly-Based}

\vspace{1mm}
\noindent\textbf{MAGIC: } This is an end-to-end malware detection method that classifies malware programs represented by Attributed Control Flow Graph (ACFG) using Deep Graph Convolutional Neural Network (DGCNN)~\cite{dgcnn2018end}. It first to convert the ACFG, which abstracts the vertices of the Control Flow Graph (CFG) as discrete numerical value vectors, extracted from malware disassembly file into a numerical vector. 
DGCNN then transforms these un-ordered ACFGs of varying sizes to tensors of fixed size and order for malware family classification. 

\vspace{1mm}
\noindent\textbf{Word2Vec+KNN: } 
This method models malware disassembly file as a malware language, extracts the opcode sequence of it as malware document and uses the Word2Vec model to generate a computational representation of such document. This work choose Word Mover’s Distance (WMD)~\cite{wmd2015word} as the measure of semantic closeness between documents for KNN classification, which computes the cost of transporting all embedded words of document $A$ to all embedded words of document $B$.

\vspace{1mm}
\noindent\textbf{MCSC: }
It first extracts opcode sequences from disassembly files and encodes them to equal length based on  SimHash~\cite{simhash2007detecting}. Then, it takes each SimHash value as a binary pixel and converts the SimHash bits to grayscale images. It trains a CNN structure modified from LeNet-5~\cite{lenet1989backpropagation} to classify these grayscale images. When training the CNN classifier, muti-hash and bilinear interpolation are used to improve the accuracy of the model, and major block selection is used to decrease the image generation time.

\subsection{Experimental Datasets}
\label{subsec:datasets}

We employ four different datasets for evaluation, i.e., BIG-15~\cite{big20152018microsoft}, Malimg~\cite{Nataraj2011malware}, MalwareBazaar and MalwareDrift, respectively. The first two datasets are adopted from prior approaches, while the last two are newly constructed for our study. BIG-15 and Malimg are commonly used by previous research work. They are open-source and well-maintained with a large number of diverse families of malware. 
MalwareBazaar is constructed due to the reason that   
currently the only public available dataset to measure all methods discussed in Sec.~\ref{subsec:techniques} is BIG-15, which was released in 2015 and could have been dated. Therefore, we build a new dataset with the latest occurred malware to eliminate the experimental bias caused by using a single dataset; 
MalwareDrift is used to evaluate the effectiveness of existing approaches when facing the scenario of concept drift. We publish all above datasets used in the study for future research in this direction.


\vspace{1mm}
\noindent \textbf{BIG-15}~\cite{big20152018microsoft} is first released in the Malware Classification Challenge on Kaggle by Microsoft in 2015. This public dataset contains 10,868 labeled PE malware samples from 9 families. The raw malware are converted to binary and disassembly files for safety concerns. 

\vspace{1mm}
\noindent \textbf{Malimg}~\cite{Nataraj2011malware} comprises of 9,435 malware collected from 25 families  in 2011. This dataset only contains the greyscale images converted from the malware binary files and is widely used by image-based malware classification approaches.

\vspace{1mm}
\noindent \textbf{MalwareBazaar} is a new dataset we constructed according to the MalwareBazaar website~\cite{malwarebazaar}, which provides free and unrestricted download services for malware samples. 
We first choose the top-6 malware families from the data released in 2020, and then  download the most recently uploaded 1,000 malware samples for each family. Then, we filter out samples that are not in PE format and further leverage Joe Security~\cite{joesecurity} and AVClass~\cite{avclass} to check the label of each sample to remove noise samples, which different website give inconsistent labels. As a result, we obtain a dataset with 3,971 PE malware samples from 6 families, which is summarized in Table~\ref{tab:new-dataset}.
To Make the dataset conform to different approaches, we further used IDA Pro~\cite{ida} to convert the malware into binary and disassembly files. 
Please note that since MalwareBazaar is constructed from the latest malware samples, it can better reflect the tendency of the latest malware, which may provide a different perspective and complement with previous datasets, such as BIG-15. 

\vspace{1mm}
\noindent \textbf{MalwareDrift} is constructed based on the conclusions by  Wadkar et al.~\cite{drift2020detecting}. Their experiments confirmed that code changes appear as sharp spikes in the $\chi^2$ timeline statistic, which quantifies the weight differences in Support Vector Machine (SVM) trained with PE malware features over the sliding time windows. Code changes can also be understood as the evolution or drift of a malware family. 
We adopt the dataset used in~\cite{drift2020detecting} as well as the corresponding $\chi^2$ timeline statistic graph, based on which we can decide the evolution time period of each malware family according to the spikes, and divide samples of each family into the  pre-drift and post-drift parts. After eliminating families without obvious sharp spikes and families of small sample size, we finally obtain the drift dataset, including 3,125 samples from 7 families, as is shown in Table~\ref{tab:drift-dataset}. 

\subsection{Experimental Setup}
\label{subsec:config}

\begin{table*}[t]
\setlength{\abovecaptionskip}{0.15cm}
\centering\footnotesize
\begin{threeparttable}
\caption{Performance of all methods on three datasets (10-fold cross-validation)}
\label{tab:three_datasets}
\begin{tabular}{c|c|c|cccc|c|ccc}
\toprule
\hline
\multirow{2}{*}{\textbf{Dataset}} & \multirow{2}{*}{\textbf{Category}} & \multirow{2}{*}{\textbf{Model}} & \multicolumn{4}{c|}{\textbf{Classification Performance (\%)}}     & \multirow{2}{*}{\textbf{Train Time (min)}} & \multicolumn{3}{c}{\textbf{Reasource Overhead}}                                                         \\ \cline{4-7} \cline{9-11} 
                                  &                                    &                                 & \textbf{$A$}     & \textbf{$P_{macro}$}     & \textbf{$R_{macro}$}     & \textbf{$F1_{macro}$}    &                                            & \multicolumn{1}{c|}{\textbf{Mem (GB)}} & \multicolumn{1}{c|}{\textbf{GPU.Mem (GB)}} & \textbf{GPU (\%)} \\ \hline
\multirow{9}{*}{BIG-15}           & \multirow{4}{*}{Image}             & ResNet-50                       & \textbf{98.42} & \textbf{96.57} & \textbf{95.68} & \textbf{96.08} & 60.13                                      & \multicolumn{1}{c|}{36.86}             & \multicolumn{1}{c|}{10.77}                 & 85                \\
                                  &                                    & VGG-16                          & 93.94          & 90.32          & 81.89          & 87.28          & 275.52                                     & \multicolumn{1}{c|}{27.65}             & \multicolumn{1}{c|}{10.90}                 & 75                \\
                                  &                                    & Inception-V3                    & 96.99          & 93.67          & 94.46          & 94.03          & 120.00                                        & \multicolumn{1}{c|}{36.86}             & \multicolumn{1}{c|}{10.77}                 & 71                \\
                                  &                                    & IMCFN                           & 97.77          & 95.93          & 94.81          & 95.13          & 52.32                                      & \multicolumn{1}{c|}{58.37}             & \multicolumn{1}{c|}{10.77}                 & 95                \\ \cline{2-11} 
                                  & \multirow{2}{*}{Binary}            & CBOW+MLP                        & \textbf{98.41} & \textbf{97.63} & \textbf{96.67} & \textbf{97.12} & 1.46                                       & \multicolumn{1}{c|}{13.31}             & \multicolumn{1}{c|}{10.44}                 & 29                \\
                                  &                                    & MalConv                         & 97.02          & 94.34          & 92.62          & 93.33          & 227.40                                     & \multicolumn{1}{c|}{259.07}            & \multicolumn{1}{c|}{10.79$\times$4}               & 37                \\ \cline{2-11} 
                                  & \multirow{3}{*}{Disassembly}       & MAGIC                           & 98.05          & \textbf{96.75} & 94.03          & 95.14          & 52.32                                      & \multicolumn{1}{c|}{13.31}             & \multicolumn{1}{c|}{9.88}                  & 73                \\
                                  &                                    & Word2Vec+KNN                    & \textbf{98.07} & 96.41          & \textbf{96.51} & \textbf{96.45} & 0.09                                       & \multicolumn{1}{c|}{1.30}              & \multicolumn{1}{c|}{-}                     & -                 \\
                                  &                                    & MCSC                            & 97.94          & 95.97          & 96.17          & 96.06          & 3.62                                       & \multicolumn{1}{c|}{3.78}              & \multicolumn{1}{c|}{10.77}                 & 33                \\ \hline
\multirow{4}{*}{Malimg}           & \multirow{4}{*}{Image}             & ResNet-50                       & 98.14          & 95.48          & 95.06          & 95.21          & 140.02                                     & \multicolumn{1}{c|}{32.77}             & \multicolumn{1}{c|}{10.77}                 & 95                \\
                                  &                                    & VGG-16                          & 99.08          & 97.71          & 97.74          & 97.73          & 58.63                                      & \multicolumn{1}{c|}{32.77}             & \multicolumn{1}{c|}{10.77}                 & 94                \\
                                  &                                    & Inception-V3                    & 97.77          & 96.29          & 93.82          & 93.98          & 273.12                                     & \multicolumn{1}{c|}{32.77}             & \multicolumn{1}{c|}{10.77}                 & 94                \\
                                  &                                    & IMCFN                           & \textbf{99.13} & \textbf{98.08} & \textbf{97.62} & \textbf{97.84} & 34.95                                      & \multicolumn{1}{c|}{43.01}             & \multicolumn{1}{c|}{10.77}                 & 96                \\ \hline
\multirow{9}{*}{MalwareBazaar}    & \multirow{4}{*}{Image}             & ResNet-50                       & 96.68          & 96.91          & 96.75          & 96.83          & 8.35                                       & \multicolumn{1}{c|}{17.41}             & \multicolumn{1}{c|}{10.77}                 & 95                \\
                                  &                                    & VGG-16                          & 96.35          & 96.58          & 96.54          & 96.56          & 43.96                                      & \multicolumn{1}{c|}{18.43}             & \multicolumn{1}{c|}{10.77}                 & 97                \\
                                  &                                    & Inception-V3                    & 95.83          & 95.67          & 95.79          & 95.73          & 6.39                                       & \multicolumn{1}{c|}{12.29}             & \multicolumn{1}{c|}{10.77}                 & 96                \\
                                  &                                    & IMCFN                           & \textbf{97.38} & \textbf{97.53} & \textbf{97.41} & \textbf{97.47} & 18.75                                      & \multicolumn{1}{c|}{22.53}             & \multicolumn{1}{c|}{10.77}                 & 92                \\ \cline{2-11} 
                                  & \multirow{2}{*}{Binary}            & CBOW+MLP                        & \textbf{97.81} & \textbf{97.92} & \textbf{98.08} & \textbf{98.00} & 0.82                                       & \multicolumn{1}{c|}{51.94}             & \multicolumn{1}{c|}{10.44}                 & 34                \\
                                  &                                    & MalConv                         & 95.92          & 96.04          & 96.43          & 96.20          & 65.40                                      & \multicolumn{1}{c|}{246.78}            & \multicolumn{1}{c|}{10.78}                 & 60                \\ \cline{2-11} 
                                  & \multirow{3}{*}{Disassembly}       & MAGIC                           & 92.82          & 88.03          & 87.36          & 87.45          & 246.00                                     & \multicolumn{1}{c|}{113.66}            & \multicolumn{1}{c|}{10.61}                 & 81                \\
                                  &                                    & Word2Vec+KNN                    & 95.64          & 93.34          & 94.29          & 93.79          & $\approx$0                                          & \multicolumn{1}{c|}{5.54}              & \multicolumn{1}{c|}{-}                     & -                 \\
                                  &                                    & MCSC                            & \textbf{96.80} & \textbf{94.97} & \textbf{94.51} & \textbf{94.70} & 1.07                                       & \multicolumn{1}{c|}{45.34}             & \multicolumn{1}{c|}{10.77}                 & 33                \\ \hline
\bottomrule
\end{tabular}
\end{threeparttable}
\vspace{-2mm}
\end{table*}
\subsubsection{Data Pre-processing}
In order to conform to the requirement of different methods and provide a consistent experiment setting, we first perform data pre-processing for the adopted datasets. (1) For image-based methods, we  first transform the malware files into the color image format as it has been shown that color images achieve better performance than greyscale images on the malware family classification task~\cite{vasan2020imcfn}. More concretely, we first transform the files into images of different width according to the file size, and then adopt the Nearest Neighbor interpolation image resize method~\cite{nearest1983comparison} to resize the image into the size of 224*224. 
It has been shown that the original texture features remain sharp with this resize method~\cite{vasan2020imcfn}. (2) For the Word2Vec+MLP method,  we follow the original settings~\cite{qiao2020malware} to remove all `0x00' and `0xcc' bytes that consecutively appear more than 5 times, and employ the Gensim library~\cite{gensim} for byte embedding. (3) For MalConv, we limit the malware file size to less than 2MB due to the reason  
that larger file sizes cause excessive GPU memory consumption~\cite{raff2017malware}. (4) For MAGIC, we adopt IDA pro~\cite{ida} to extract the Attributed Control Flow Graph (ACFG) for all malware files into documents. (5) For the Word2Vec+KNN method, we use the 
sed stream editor~\cite{sed} written in shell script to convert all assembly code to x86-64 opcode sequences. (6) For MCSC method, we apply SimHash-768~\cite{hash} to convert the Opcode sequence into a SimHash sequence, and then convert it to a black and white image and utilize the bilinear interpolation to uniformly zoom the original image to the size of 32x32. 

\subsubsection{Configurations}
In order to provide a fair comparison, we adopt the default hyper-parameter settings for the methods if they are released in the original paper~\cite{yan2019classifying, awad2018modeling, ni2018malware}, which are said to achieve the best performance of the corresponding  methods. We only extract the opcode sequences in the Word2Vec+KNN approach~\cite{awad2018modeling} for the sake of time. 
For the other methods which do not report the hyper-parameters leading to the best performance, we perform an intensive manual tuning process and employ the settings that achieve the best performance. Particularly, we apply the early stopping mechanism to fairly compare the learning efficiency for different methods. 
Due to the space limit, we do not report the detailed experiment settings in the paper, the information is available in our open-source repository.

As transfer learning is usually adopted for image-based methods in the malware family classification task to enhance the performance. We also evaluate the effect of transfer learning in our study. Following the standard process, we fix the structure (i.e., the number of layers and neurons per layer) of each model. We first perform pre-training with the ImageNet dataset~\cite{imagenet2009imagenet} and then fine tune the models with our malware datasets in image format. To explore the effect of dataset size on performance, we use 10\%, 50\%, 80\% and 100\%  of the malware data, respectively, for fine-tuning.


Particularly, to evaluate the performance of different methods in  the concept drift scenario, we use the pre-drift samples to train a model following a standard data partition of 8:2 for training and testing, and report the performance of models on the pre-drift data. Then, we load the trained model and test it on the post-drift data. 

Following the standard paradigm, we apply 10-fold cross-validation in our experimental comparison among different methods on different datasets, and 5-fold cross-validation on the concept drift experiment. We utilize the Macro Average Metrics of Precision ($P_{macro}$), Recall ($R_{macro}$) and F1-Score ($F1_{macro}$) to measure the multi-classification performance of a model on multiple malware families. 
Suppose there are $N$ malware families (with a total of $S$ samples) and we conduct $M$-fold cross-validation. First we calculate the total Precision ($P_n$), Recall ($R_n$) and F1-Score ($F1_n$) of $M$-fold cross-validation for each family class $n$ ($1 \leq n \leq N$). 
The formulas used to calculate the metrics are shown in fomulas~\ref{eq:P}-\ref{eq:F}, where \textbf{$TP_{mn}$}, \textbf{$FP_{mn}$} and \textbf{$FN_{mn}$} represent the true positive, false positive and false negative of malware family $n$ in the $m$-fold. 

\begin{equation}
\small{
\centering
\label{eq:P}
P_{n}=(\sum_{m=1}^M TP_{mn}) / (\sum_{m=1}^M (TP_{mn}+FP_{mn}))
}
\end{equation}

\begin{equation}
\small{
\centering
\label{eq:R}
R_{n}=(\sum_{m=1}^M TP_{mn}) / (\sum_{m=1}^M (TP_{mn}+FN_{mn}))
}
\end{equation}

\begin{equation}
\small{
\centering
\label{eq:F}
F1_{n}=(2*P_{n}*R_{n}) / (P_{n}+R_{n})
}
\end{equation}

Based on the per-family precision ($P_n$), recall ($R_n$) and F1-score ($F_n$) ($1 \leq n \leq N$) computed with the $M$-fold data, the Macro Average of Precision, Recall and F1-score for multiple families are  calculated with formula~\ref{eq:PM}-\ref{eq:FM}. We also compute the  accuracy with formula~\ref{eq:A}. 

\begin{equation}
\small{
\centering
\label{eq:A}
A=(\sum_{n=1}^N \sum_{m=1}^M TP_{mn}) / S
}
\end{equation}

\begin{equation}
\small{
\centering
\label{eq:PM}
P_{macro}=(\sum_{n=1}^N P_n) / N
}
\end{equation}

\begin{equation}
\small{
\centering
\label{eq:RM}
R_{macro}=(\sum_{n=1}^N R_n) / N
}
\end{equation}

\begin{equation}
\small{
\centering
\label{eq:FM}
F1_{macro}=(\sum_{n=1}^N F1_n) / N
}
\end{equation}


\subsubsection{Executing Environment}
All of our experiments are conducted on a server with 2 Intel Xeon Platinum 8260 CPU @2.30GHz, 8 Nvidia GeForce RTX2080 Ti GPU (11GB memory), and 512GB RAM.

\section{Experimental Result and Analysis}
\label{sec:results}
\begin{table*}[t]
\setlength{\abovecaptionskip}{0.15cm}
\centering\footnotesize
\begin{threeparttable}
\caption{Performance of transfer learning for image-based approaches on three datasets (10-fold cross-validation) }
\label{tab:transfer_three_datasets}
\begin{tabular}{l|c|c|cccc|c|ccc}
\toprule
\hline
\multirow{2}{*}{\textbf{Dataset}} & \multirow{2}{*}{\textbf{Model}} & \multirow{2}{*}{\textbf{Training Strategy}} & \multicolumn{4}{c|}{\textbf{Classification Performance (\%)}}     & \multirow{2}{*}{\textbf{Train Time (min)}} & \multicolumn{3}{c}{\textbf{Reasource Overhead}}                                                         \\ \cline{4-7} \cline{9-11} 
                                  &                                 &                                             & \textbf{$A$}     & \textbf{$P_{macro}$}     & \textbf{$R_{macro}$}     & \textbf{$F1_{macro}$}    &                                            & \multicolumn{1}{c|}{\textbf{Mem (GB)}} & \multicolumn{1}{c|}{\textbf{GPU.Mem (GB)}} & \textbf{GPU (\%)} \\ \hline
\multirow{16}{*}{BIG-15}          & \multirow{4}{*}{ResNet-50}      & 10\% TL                                     & 94.76          & 92.74          & 82.69          & 86.77          & 3.97                                       & \multicolumn{1}{c|}{22.53}             & \multicolumn{1}{c|}{10.56}                 & 3                 \\
                                  &                                 & 50\% TL                                     & 96.68          & 95.35          & 85.22          & 91.50          & 24.78                                      & \multicolumn{1}{c|}{22.53}             & \multicolumn{1}{c|}{10.56}                 & 3                 \\
                                  &                                 & 80\% TL                                     & 96.90          & 94.87          & 85.53          & 92.50          & 37.02                                      & \multicolumn{1}{c|}{23.55}             & \multicolumn{1}{c|}{10.56}                 & 3                 \\
                                  &                                 & 100\% TL                                    & \textbf{97.05} & \textbf{95.64} & \textbf{93.60} & \textbf{94.48} & 43.62                                      & \multicolumn{1}{c|}{21.50}             & \multicolumn{1}{c|}{10.56}                 & 4                 \\ \cline{2-11} 
                                  & \multirow{4}{*}{VGG-16}         & 10\% TL                                     & 89.56          & 78.74          & 76.73          & 77.59          & 0.42                                       & \multicolumn{1}{c|}{19.46}             & \multicolumn{1}{c|}{10.44}                 & 3                 \\
                                  &                                 & 50\% TL                                     & 95.21          & 89.02          & 83.77          & 87.21          & 11.76                                      & \multicolumn{1}{c|}{21.50}             & \multicolumn{1}{c|}{10.44}                 & 44                \\
                                  &                                 & 80\% TL                                     & 96.62          & 92.00          & 85.20          & 90.41          & 8.04                                       & \multicolumn{1}{c|}{22.53}             & \multicolumn{1}{c|}{10.44}                 & 71                \\
                                  &                                 & 100\% TL                                    & \textbf{96.78} & \textbf{93.36} & \textbf{91.78} & \textbf{92.46} & 9.72                                       & \multicolumn{1}{c|}{23.55}             & \multicolumn{1}{c|}{10.44}                 & 70                \\ \cline{2-11} 
                                  & \multirow{4}{*}{Inception-V3}   & 10\% TL                                     & 92.28          & 91.46          & 80.51          & 80.94          & 3.27                                       & \multicolumn{1}{c|}{18.43}             & \multicolumn{1}{c|}{10.56}                 & 4                 \\
                                  &                                 & 50\% TL                                     & 95.10          & 91.17          & 86.76          & 86.87          & 12.78                                      & \multicolumn{1}{c|}{19.46}             & \multicolumn{1}{c|}{10.56}                 & 8                 \\
                                  &                                 & 80\% TL                                     & 95.31          & 92.52          & 88.54          & 88.64          & 19.02                                      & \multicolumn{1}{c|}{19.46}             & \multicolumn{1}{c|}{10.56}                 & 8                 \\
                                  &                                 & 100\% TL                                    & \textbf{95.56} & \textbf{93.48} & \textbf{89.64} & \textbf{89.76} & 22.92                                      & \multicolumn{1}{c|}{21.50}             & \multicolumn{1}{c|}{10.56}                 & 17                \\ \cline{2-11} 
                                  & \multirow{4}{*}{IMCFN}          & 10\% TL                                     & 94.81          & 94.13          & 84.04          & 85.21          & 4.52                                       & \multicolumn{1}{c|}{20.48}             & \multicolumn{1}{c|}{10.44}                 & 25                \\
                                  &                                 & 50\% TL                                     & 97.38          & 95.25          & 91.90          & 93.15          & 14.13                                      & \multicolumn{1}{c|}{22.53}             & \multicolumn{1}{c|}{10.44}                 & 34                \\
                                  &                                 & 80\% TL                                     & 97.81          & 96.09          & 93.69          & 94.69          & 21.04                                      & \multicolumn{1}{c|}{28.67}             & \multicolumn{1}{c|}{10.44}                 & 40                \\
                                  &                                 & 100\% TL                                    & \textbf{98.05} & \textbf{96.18} & \textbf{94.37} & \textbf{95.18} & 21.41                                      & \multicolumn{1}{c|}{32.77}             & \multicolumn{1}{c|}{10.44}                 & 40                \\ \hline
\multirow{16}{*}{Malimg}          & \multirow{4}{*}{ResNet-50}      & 10\% TL                                     & 97.34          & 93.79          & 92.77          & 93.21          & 5.29                                       & \multicolumn{1}{c|}{19.46}             & \multicolumn{1}{c|}{10.56}                 & 7                 \\
                                  &                                 & 50\% TL                                     & 98.02          & 95.61          & 94.46          & 94.93          & 29.61                                      & \multicolumn{1}{c|}{19.46}             & \multicolumn{1}{c|}{10.56}                 & 5                 \\
                                  &                                 & 80\% TL                                     & \textbf{98.59} & \textbf{96.37} & \textbf{96.23} & \textbf{96.29} & 44.65                                      & \multicolumn{1}{c|}{19.46}             & \multicolumn{1}{c|}{10.56}                 & 8                 \\
                                  &                                 & 100\% TL                                    & 98.49          & 96.33          & 95.88          & 96.10          & 70.80                                      & \multicolumn{1}{c|}{20.48}             & \multicolumn{1}{c|}{10.56}                 & 7                 \\ \cline{2-11} 
                                  & \multirow{4}{*}{VGG-16}         & 10\% TL                                     & 96.37          & 90.05          & 89.24          & 89.50          & 2.40                                       & \multicolumn{1}{c|}{17.41}             & \multicolumn{1}{c|}{10.44}                 & 7                 \\
                                  &                                 & 50\% TL                                     & 97.24          & 91.29          & 91.64          & 91.41          & 7.32                                       & \multicolumn{1}{c|}{16.38}             & \multicolumn{1}{c|}{10.44}                 & 5                 \\
                                  &                                 & 80\% TL                                     & 98.58          & 96.35          & 96.23          & 96.28          & 11.46                                      & \multicolumn{1}{c|}{17.41}             & \multicolumn{1}{c|}{10.44}                 & 20                \\
                                  &                                 & 100\% TL                                    & \textbf{98.63} & \textbf{96.47} & \textbf{96.40} & \textbf{96.42} & 14.16                                      & \multicolumn{1}{c|}{19.46}             & \multicolumn{1}{c|}{10.44}                 & 52                \\ \cline{2-11} 
                                  & \multirow{4}{*}{Inception-V3}   & 10\% TL                                     & 93.82          & 87.07          & 83.93          & 85.05          & 3.93                                       & \multicolumn{1}{c|}{16.38}             & \multicolumn{1}{c|}{10.56}                 & 6                 \\
                                  &                                 & 50\% TL                                     & 95.75          & 89.23          & 88.63          & 88.86          & 11.04                                      & \multicolumn{1}{c|}{17.41}             & \multicolumn{1}{c|}{10.56}                 & 2                 \\
                                  &                                 & 80\% TL                                     & 96.17          & 90.15          & 89.61          & 89.84          & 16.32                                      & \multicolumn{1}{c|}{20.48}             & \multicolumn{1}{c|}{10.56}                 & 12                \\
                                  &                                 & 100\% TL                                    & \textbf{96.43} & \textbf{90.62} & \textbf{90.28} & \textbf{90.41} & 18.96                                      & \multicolumn{1}{c|}{18.43}             & \multicolumn{1}{c|}{10.56}                 & 7                 \\ \cline{2-11} 
                                  & \multirow{4}{*}{IMCFN}          & 10\% TL                                     & 91.48          & 74.18          & 74.05          & 73.35          & 4.63                                       & \multicolumn{1}{c|}{18.43}             & \multicolumn{1}{c|}{10.44}                 & 26                \\
                                  &                                 & 50\% TL                                     & 96.67          & 90.70          & 90.06          & 90.27          & 12.66                                      & \multicolumn{1}{c|}{22.52}             & \multicolumn{1}{c|}{10.44}                 & 35                \\
                                  &                                 & 80\% TL                                     & 96.96          & 91.17          & 90.85          & 90.93          & 18.41                                      & \multicolumn{1}{c|}{22.53}             & \multicolumn{1}{c|}{10.44}                 & 39                \\
                                  &                                 & 100\% TL                                    & \textbf{97.13} & \textbf{91.29} & \textbf{91.33} & \textbf{91.25} & 18.55                                      & \multicolumn{1}{c|}{23.55}             & \multicolumn{1}{c|}{10.44}                 & 40                \\ \hline
\multirow{16}{*}{MalwareBazaar}   & \multirow{4}{*}{ResNet-50}      & 10\% TL                                     & 90.03          & 89.48          & 90.99          & 90.09          & 4.07                                       & \multicolumn{1}{c|}{9.14}              & \multicolumn{1}{c|}{10.56}                 & 6                 \\
                                  &                                 & 50\% TL                                     & 93.98          & 94.28          & 94.57          & 94.38          & 7.23                                       & \multicolumn{1}{c|}{9.13}              & \multicolumn{1}{c|}{10.56}                 & 7                 \\
                                  &                                 & 80\% TL                                     & 94.90          & 95.21          & 95.34          & 95.25          & 9.11                                       & \multicolumn{1}{c|}{9.16}              & \multicolumn{1}{c|}{10.56}                 & 5                 \\
                                  &                                 & 100\% TL                                    & \textbf{95.30} & \textbf{95.66} & \textbf{95.68} & \textbf{95.65} & 10.01                                      & \multicolumn{1}{c|}{9.19}              & \multicolumn{1}{c|}{10.56}                 & 6                 \\ \cline{2-11} 
                                  & \multirow{4}{*}{VGG-16}         & 10\% TL                                     & 83.55          & 86.50          & 76.12          & 76.80          & 1.16                                       & \multicolumn{1}{c|}{8.17}              & \multicolumn{1}{c|}{10.44}                 & 56                \\
                                  &                                 & 50\% TL                                     & 92.35          & 92.75          & 92.54          & 92.51          & 2.19                                       & \multicolumn{1}{c|}{9.34}              & \multicolumn{1}{c|}{10.44}                 & 52                \\
                                  &                                 & 80\% TL                                     & 93.85          & 93.92          & 94.37          & 94.06          & 2.98                                       & \multicolumn{1}{c|}{10.24}             & \multicolumn{1}{c|}{10.44}                 & 55                \\
                                  &                                 & 100\% TL                                    & \textbf{94.43} & \textbf{94.43} & \textbf{94.90} & \textbf{94.60} & 3.94                                       & \multicolumn{1}{c|}{11.26}             & \multicolumn{1}{c|}{10.44}                 & 52                \\ \cline{2-11} 
                                  & \multirow{4}{*}{Inception-V3}   & 10\% TL                                     & 90.28          & 90.87          & 90.99          & 90.87          & 0.79                                       & \multicolumn{1}{c|}{7.83}              & \multicolumn{1}{c|}{10.56}                 & 7                 \\
                                  &                                 & 50\% TL                                     & 93.75          & 94.31          & 94.20          & 94.24          & 3.77                                       & \multicolumn{1}{c|}{8.13}              & \multicolumn{1}{c|}{10.56}                 & 8                 \\
                                  &                                 & 80\% TL                                     & 94.48          & 94.96          & 94.87          & 94.90          & 6.37                                       & \multicolumn{1}{c|}{8.70}              & \multicolumn{1}{c|}{10.56}                 & 6                 \\
                                  &                                 & 100\% TL                                    & \textbf{94.80} & \textbf{95.15} & \textbf{95.17} & \textbf{95.15} & 8.47                                       & \multicolumn{1}{c|}{8.47}              & \multicolumn{1}{c|}{10.56}                 & 7                 \\ \cline{2-11} 
                                  & \multirow{4}{*}{IMCFN}          & 10\% TL                                     & 91.30          & 91.72          & 91.27          & 91.35          & 2.14                                       & \multicolumn{1}{c|}{9.39}              & \multicolumn{1}{c|}{10.44}                 & 27                \\
                                  &                                 & 50\% TL                                     & 94.95          & 94.99          & 95.33          & 95.10          & 5.75                                       & \multicolumn{1}{c|}{10.24}             & \multicolumn{1}{c|}{10.44}                 & 37                \\
                                  &                                 & 80\% TL                                     & 95.55          & 95.59          & 95.86          & 95.68          & 6.25                                       & \multicolumn{1}{c|}{11.26}             & \multicolumn{1}{c|}{10.44}                 & 45                \\
                                  &                                 & 100\% TL                                    & \textbf{95.73} & \textbf{95.73} & \textbf{96.02} & \textbf{95.84} & 7.76                                       & \multicolumn{1}{c|}{11.26}             & \multicolumn{1}{c|}{10.44}                 & 37                \\ \hline
\bottomrule
\end{tabular}
\end{threeparttable}
\vspace{-2mm}
\end{table*}

\subsection{RQ1: Performance Comparison}
\label{subsec:RQ1}



Table~\ref{tab:three_datasets} shows the experimental results of all adopted models on the three studied datasets. Particularly, Malimg was only used for image-based models as the original malware file is unavailable.

From the table we can see that the method with CBOW+MLP model achieves the best performance in terms of F1-score compared with all the others on both BIG-15 and MalwareBazaar datasets, indicating the generalizability of it to conform different datasets. On the other hand, when considering the performance of different categories, there is no individual category that always outperforms the others. For example, for binary-based methods,  though the CBOW+MLP model performs best, the MalConv model is not always better than the other methods. Base on this result, we can conclude that the data format of malware should not be a critical factor that impacts the classification performance, which is more likely decided by the model itself. 
Besides, by comparing the experimental results across different datasets for each model, most of models can achieve stable performance except VGG-16 and MAGIC, whose F1-scores vary greatly on the BIG-15 and MalwareBazzar datasets. We further analyze the results and observe that imbalanced data is a major reason for VGG-16. For example, it performs much worse on BIG-15 compared with other datasets, because there are only 42 samples belonging to the Simda family, 
which only accounts for 1.43\%-10.55\% of the other families and less than 0.4\% of the total dataset. Therefore, VGG-16 can hardly learn the deep semantics from such limited samples as it has the largest number of parameters (see Table~\ref{tab:model-deployment}). While for MAGIC, the reason is that it has large runtime GPU memory consumption, especially when the input file size is large. For BIG-15, the largest disassembly file is 140MB, while for MalwareBazaar, the file size can exceed 1GB. We can only set the batch size to 1 when processing the MalwareBazaar dataset due to the limitations of our GPU memory. The small batch size could affect the performance of the method. 

\begin{table*}[t]
\setlength{\abovecaptionskip}{0.15cm}
\centering
\begin{threeparttable}
\small
\caption{The resource consumption of various methods in runtime}
\label{tab:model-deployment}
\begin{tabular}{c|c|c|c|cc|cc|c}
\toprule
\hline
\multirow{3}{*}{Category}    & \multirow{3}{*}{Model} & \multicolumn{2}{c|}{Time}                                        & \multicolumn{4}{c|}{Saved Model Size (MB)}                                      & \multicolumn{1}{l}{\multirow{3}{*}{Runtime memory (MB)}} \\ \cline{3-8}
                             &                        & \multirow{2}{*}{Pre-process (s)} & \multirow{2}{*}{Predict (ms)} & \multicolumn{2}{c|}{Save Weights Only} & \multicolumn{2}{c|}{Save Entire Model} & \multicolumn{1}{l}{}                                       \\ \cline{5-8}
                             &                        &                                  &                               & From Scratch         & Transfer        & From Scratch         & Transfer        & \multicolumn{1}{l}{}                                       \\ \hline
\multirow{4}{*}{Image}       & ResNet-50              & \multirow{4}{*}{0.7}             & 2.62                          & 90.27                & 90.45           & 270.51               & 180.94          & 2359.50                                                    \\
                             & VGG-16                 &                                  & 2.21                          & 512.36               & 512.36          & 1024.00              & 1024.73         & 2312.17                                                    \\
                             & Inception-V3           &                                  & 2.03                          & 83.91                & 83.91           & 250.77               & 168.04          & 2369.00                                                    \\
                             & IMCFN                  &                                  & 2.23                          & 268.28               & 268.28          & 536.57               & 536.57          & 2299.00                                                    \\ \hline
\multirow{2}{*}{Binary}      & CBOW+MLP               & 1.09                             & 5441.02                       & \multicolumn{2}{c|}{129.03}            & \multicolumn{2}{c|}{258.07}            & 1406.67                                                    \\
                             & MalConv                & 0.28                             & 14.02                         & \multicolumn{2}{c|}{3.96}              & \multicolumn{2}{c|}{7.91}              & 2378.50                                                    \\ \hline
\multirow{3}{*}{Disassembly} & MAGIC                  & 17.78                            & 23.63                         & \multicolumn{2}{c|}{412.14}            & \multicolumn{2}{c|}{1239.04}           & 3898.83                                                    \\
                             & Word2Vec+KNN           & 17.72                            & 95243.31$^\star$                     & \multicolumn{2}{c|}{-}                & \multicolumn{2}{c|}{-}                & -                                                         \\
                             & MCSC                   & 4.49                             & 3477.77                       & \multicolumn{2}{c|}{3.63}              & \multicolumn{2}{c|}{7.27}              & 2275.67                                                    \\ \hline
\bottomrule
\end{tabular}
\footnotesize
 We report the average time taken to process and predict one sample; $^\star$: Contains the time to calculate 6893 document distances;$-$: KNN cannot generate a model that can be saved;
The sizes reported on transfer learning contain the parameters of the pre-trained feature extraction layers, the fine-tuned classification layer and the model structure (for the entire model setting only).
\end{threeparttable}
\vspace{-3mm}
\end{table*}

Considering each individual category, though IMCFN in general achieves the best performance (average F1-score: 96.81\%) compared with the other image-based methods, there is no evidence 
that it significantly outperforms the others. Similar to CBOW+MLP (average F1-score: 97.56\%) and MCSC (average F1-score: 95.38\%) for the binary-based and disassembly-based methods. The overall performance within each category is relatively close to each other.

\finding{The binary-based model CBOW+MLP performs the best across different datasets among all methods, while no individual category significantly outperforms the others. }

\jjcmt{The F1-score for image-based methods ranges from 87.28\% to 97.84\% on all the studied datasets, while it ranges from 93.33\% to 98.00\% for binary-based and from 87.45\% to 96.45\% for disassembly-based methods. Therefore, in general binary-based methods have the best stability.}

\jjcmt{The F1-score for the image-based methods, which train the models from scratch, on BIG-15, Malimg and MalwareBazzar datasets are 93.13\%, 96.17\% and 96.65\%, respectively.
The F1-score for the binary-based and disassembly-based methods on BIG-15 dataset are 95.23\% and 97.1\%, and the values on the MalwareBazzar dataset are 95.55\% and 91.98\%, respectively. 
We can observe that the binary-based approaches achieve consistent good performance as compared with  the other kinds of methods on both BIG-15 and MalwareBazzar datasets.  
The disassembly-based methods show unstable performance, which is the best on the BIG-15 dataset, yet the worst on the MalwareBazzar dataset. \ls{The potential reason for the unstable performance of the disassemble-based methods.} 
The image-based methods show a performance in-between the other two kinds of methods. }

\jjcmt{Particularly, VGG-16 achieves a  F1-score of 87.28\% on the BIG-15 dataset, which is the worst among all methods. However, VGG-16 shows good performances on the other two datasets. 
By a close look into the classification results of the BIG-15 dataset, we find that the decrease of performance is mainly caused by the Simda family, which has only 42 samples and is only 1.43\%-10.55\% of the size of other families. Since VGG-16 has the largest number of parameters among all models, as is shown in Table~\ref{tab:model-deployment}, it requires more samples to fully train the model parameters. Therefore, VGG-16 does not learn enough feature of the Simda family, which further affects the overall performance. 
%
The good performance of IMCFN is related to its model structure. IMCFN has nearly 85M parameters, while ResNet-50 and Inception-V3 only have less than 25M parameters. Under training from scratch strategy, a large number of parameters allow it to fit the malware image very well. Besides, compared to VGG-16, the addition of a dropout layer and fewer nodes in fully connected layers can effectively prevent overfitting.}

\begin{figure}[t]
    \centering
    \vbox to 2.2cm{
    \includegraphics[width=.48\textwidth]{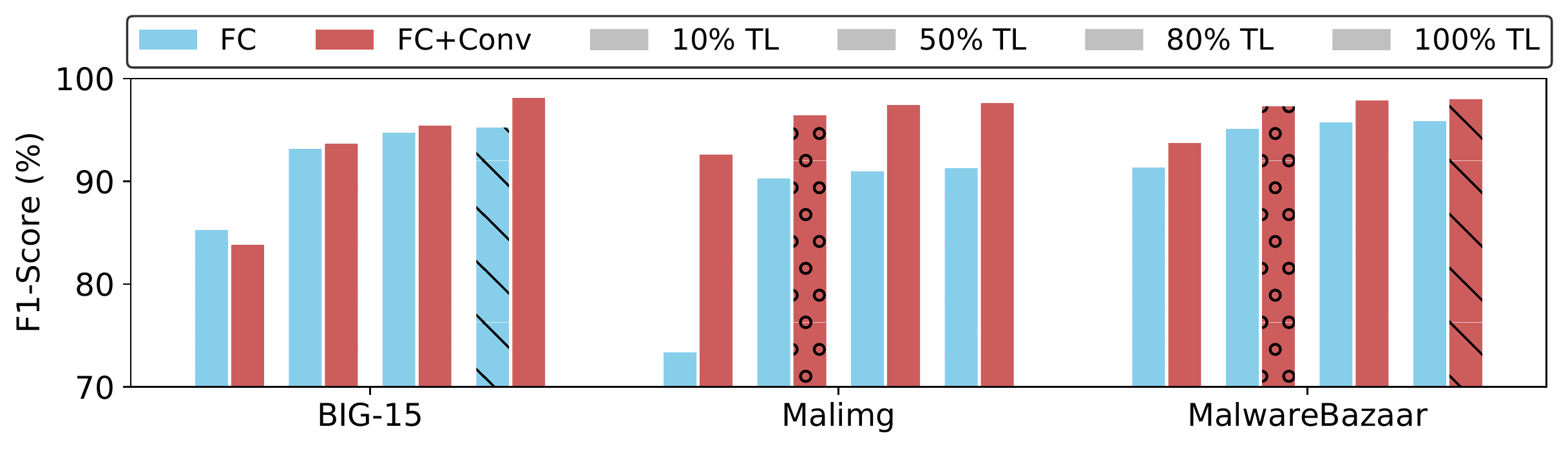}
    }
    \caption{The F1-score of only open fully-connected layers (FC) vs. the aggressive strategy (FC+Conv) of IMCFN. } 
    \label{fig:IMCFNtransfer}
    \vspace{-5mm}
\end{figure}

It has been demonstrated above that insufficient training data for large-scale networks (e.g., VGG-16) may cause big performance drop.   Transfer learning has been widely adopted in the image processing domain to tackle similar issues and obtained good performance, and thus we also investigate its impact on PE malware classification task in this work. We utilize the pre-trained model on ImageNet~\cite{imagenet2009imagenet} and use the corresponding malware image data to fine tune the last fully-connected layer. 
The experimental results are shown in Table~\ref{tab:transfer_three_datasets}. Specifically, VGG-16, which did not perform well due to insufficient training data, was significantly improved from 87.28\% to 92.46\% on BIG-15 through transfer learning. However, according to the results, the impact of transfer learning was still limited and the performance may largely drop for some cases, e.g., IMCFN. 
As a result, we further employ a more aggressive strategy that permits to update the connection weights in the last representation layer during fine-tuning. Figure~\ref{fig:IMCFNtransfer} shows results of the IMCFN model. From the figure, we can observe that opening both the fully connected layer and the convolution layers in general leads to better performance, especially on Malimg. We also measure the cost of fine-tuning the last fully-connected layer only and the aggressive strategy, where we find that the latter causes 1.5x-3x training time (less than 1 hour) compared with the former, yet gains an average of 4.14\% performance increase in terms of F1-score. More importantly, the aggressive strategy exactly advances  the classification performance, in terms of F1-score , by  0.51\%-2.94\% compared with training from scratch, except the Malimg dataset: 97.59\% (aggressive) and 97.84\% (from scratch).

\jjcmt{Transfer learning, which is widely adopted in the image processing domain, does not show consistent performance gain in most models on all datasets. In particular, VGG-16 shows 5.18\% F1-score increase and IMCFN shows 0.05\% F1-score increase on the BIG-15 dataset, and ResNet-50 shows 0.89\% F1-score increase on the Malimage dataset, with all data used for transfer learning. On all the other cases, transfer learning shows no performance gain. }

\jjcmt{Since the current transfer learning setting is fine-tuning the network parameters with only the fully connected layers opened for update with the malware image data, we would like to observe the performance of transfer learning when we open some layers in the representation part of the network. The results are shown in Figure~\ref{fig:IMCFNtransfer}. We can observe that opening both the fully connected layer and the convolution layers in general shows better performance than only open the fully connected layer for fine-turning. The performance gain is rather large on the Malimg dataset. We also measure on the training cost of those two settings, and opening both the fully connected layer and the convolution layers for fine-tuning consumes 1.5-3 times training time compared with only open the fully connected layer for fine-tuning, which is within 1 hour and is acceptable as a one-time cost. } 
The results indicate that converted malware images are different, in terms of image features, from typical images in ImageNet. Therefore, the features trained with ImageNet are not directly applicable to malware images, and thus we need to open the feature extraction layers (i.e., the convolution layers in the models we studied) in the fine-tuning process for a better performance. 

\finding{Transfer learning can potentially further improve the effectiveness of image-based PE malware classification methods, and  internal feature extraction layers of the model should be opened for fine-tuning to achieve better transfer-learning performance.}

Besides effectiveness, another important factor that impacts the practicality of PE malware classification methods in industrial scenarios is the efficiency and requirement of hardware environment since some methods may be required to work on resource-critical devices, such as a network gateway device with limited computing resources.  
Table~\ref{tab:three_datasets} and Table~\ref{tab:transfer_three_datasets} show the time and consumption of resources in detail for each method during training of the corresponding models, and the average runtime resource consumption of each method is shown in Table~\ref{tab:model-deployment}. 
From the tables we can see that image-based methods are CPU intensive compared to the other categories, while transfer learning is a possible solution as it can help to reduce 28.49\%$\sim$96.47\% training time. In addition, according to  Table~\ref{tab:model-deployment}, image-based methods take very small pre-process and prediction overhead. The model sizes are relatively larger than the other methods. Disassembly-based methods take relatively longer pre-processing time, due to the time-consuming disassemble process. Particularly, the prediction time for the Word2Vec+KNN method is much longer, because it requires computing the distances between the given input data with all samples in the training set, all such factors indeed 
restrict the application of existing methods in the industry, and should be taken into consideration in future research.


\jjcmt{From the perspective of efficiency, the training resource consumption of each method on each dataset is shown in Table~\ref{tab:BIG15}-Table~\ref{tab:MalwareBazaar}, respectively, and the average runtime resource consumption of each method is shown in Table~\ref{tab:model-deployment}. We can observe that the training time of different methods vary from a few minutes to more than four hours. The memory consumption varies from a few GB to 259 GB for different models and different dataset. The GPU memory assumption is almost all around 10-11 GB due to the reasons that the GPU in our server has 11 GB memory and the platform attempts to request as many memory as possible. The image-based methods tends to consume more CPU as compared to other kinds approaches. In general, transfer learning reduces the training time by a factor of 28.49\% $-$ 96.47\%. As the resource consumption is one-time effort, the different approaches do not show significant difference if a standard server is available. }

\jjcmt{We can also observe from Table~\ref{tab:model-deployment} that image-based methods takes very small pre-process and prediction overhead. The model sizes are relatively larger than the other methods. Disassembly-based methods take relatively longer pre-processing time, due to the  time-consuming disassemble process. The prediction time for the Word2Vec+KNN method is long, due to the reason that it requires computing the distances between the given input data with all samples in the training set. }

\jjcmt{
\finding{In general, IMCFN and CBOW+MLP consistently show the best performance across different datasets.  Feature extraction layers should be opened for fine-turning to achieve better transfer-learning performance.}
}

\subsection{RQ2: Performance under Concept Drift}
\label{subsec:RQ2}

\begin{table}[t]
\setlength{\abovecaptionskip}{0.15cm}
\begin{threeparttable}
\footnotesize
\caption{Impact of Concept Drift on Performance}
\label{tab:drity whole}
\begin{tabular}{c|c|c|cccc}
\toprule
\hline
\multirow{2}{*}{Category}    & \multirow{2}{*}{Model}                             & \multicolumn{1}{l|}{\multirow{2}{*}{Drift Part}} & \multicolumn{4}{c}{Classification Performance (\%)} \\ \cline{4-7} 
                             &                                                    & \multicolumn{1}{l|}{}                            & A         & P        & R        & F                 \\ \hline
\multirow{3}{*}{Image}       & \multirow{3}{*}{IMCFN}                             & pre-drift                                        & 85.21     & 85.23    & 83.00    & \textbf{83.91}    \\
                             &                                                    & post-drift                                       & 49.90     & 52.74    & 44.42    & 42.10             \\
                             &                                                    & decrease                                         & \cellcolor{red!35}35.31     & \cellcolor{red!32}32.49    & \cellcolor{red!39}38.58    & \cellcolor{red!42}41.81             \\ \hline
\multirow{6}{*}{Binary}      & \multirow{3}{*}{CBOW+MLP}                          & pre-drift                                        & 81.69     & 83.36    & 78.58    & 80.50             \\
                             &                                                    & post-drift                                       & 17.44     & 11.52    & 14.45    & 10.88             \\
                             &                                                    & decrease                                         & \cellcolor{red!64}64.25     & \cellcolor{red!72}71.84    & \cellcolor{red!64}64.13    & \cellcolor{red!70}\textbf{69.62}    \\ \cline{2-7} 
                             & \multirow{3}{*}{MalConv}                           & pre-drift                                        & 77.43     & 78.49    & 74.97    & 76.19             \\
                             &                                                    & post-drift                                       & 46.50     & 39.19    & 39.49    & 35.51             \\
                             &                                                    & decrease                                         & \cellcolor{red!31}30.93     & \cellcolor{red!39}39.30    & \cellcolor{red!35}35.48    & \cellcolor{red!41}40.68             \\ \hline
\multirow{9}{*}{Disassembly} & \multirow{3}{*}{MAGIC}                             & pre-drift                                        & 80.85     & 81.29    & 77.77    & 79.30             \\
                             &                                                    & post-drift                                       & 42.15     & 34.69    & 33.07    & 28.30             \\
                             &                                                    & decrease                                         & \cellcolor{red!39}38.70     & \cellcolor{red!47}46.60    & \cellcolor{red!45}44.70    & \cellcolor{red!51}51.00             \\ \cline{2-7} 
                             & \multicolumn{1}{l|}{\multirow{3}{*}{Word2Vec+KNN}} & pre-drift                                        & 81.85     & 80.77    & 79.79    & 80.12             \\
                             & \multicolumn{1}{l|}{}                              & post-drift                                       & 49.18     & 48.73    & 51.80    & \textbf{43.87}    \\
                             & \multicolumn{1}{l|}{}                              & decrease                                         & \cellcolor{red!33}32.67     & \cellcolor{red!32}32.04    & \cellcolor{red!28}27.99    & \cellcolor{red!36}36.25             \\ \cline{2-7} 
                             & \multirow{3}{*}{MCSC}                              & pre-drift                                        & 74.31     & 69.44    & 73.49    & 70.89             \\
                             &                                                    & post-drift                                       & 50.58     & 46.97    & 48.25    & 43.82             \\
                             &                                                    & decrease                                         & \cellcolor{red!24}23.73     & \cellcolor{red!22}22.47    & \cellcolor{red!25}25.24    & \cellcolor{red!27}\textbf{27.07}    \\ \hline
\bottomrule
\end{tabular}
\end{threeparttable}
\end{table}

Table~\ref{tab:drity whole} shows the evaluation results of different methods in the concept drift scenario. In particular, we choose the IMCFN as a representation of the image-based approaches as it shows the best overall performance (highest average F1-score) from the previous research question. In the table, we visualize the impact of concept drift for different methods with gradient color, the deeper the red color is, the higher the score drops. 

From Table~\ref{tab:drity whole}, we can observe that all existing methods suffer from a large performance drop while confronting the concept drift in real industry scenarios, where the reduction of F1-score is up to 27.07\%-69.62\%. The F1-scores for all methods on the post-drift dataset are no more than 45\%, reflecting that existing methods fail to consider the scenario of concept drift and there is still much improvement space. By comparing the results on concept-drift and non-concept-drift datasets, we can see that it is vital to take concept drift scenario into consideration for method evaluation.

\finding{All existing methods suffer from poor performance when facing concept drift from real industrial scenarios, and thus they should be seriously considered when evaluating PE malware family classification methods.}

According to the results, the stability of different methods also varies greatly. The method of CBOW+MLP, though performs best when used in the common machine learning scenario, has the sharpest decrease in the concept drift scenario, which is mainly due to its simple structure. On the contrary, MCSC and Word2Vec+KNN show the least decrease ratios. The reason is that both of them extract Opcode sequences from disassembly files, and focus on the local context connection of Opcode sequences which tends to be retained during malware evolving and thus those methods show stable performance on concept drift. An interesting finding is that though Word2Vec+KNN  also has simple structures, it performs better than CBOW+MLP under concept drift, which credits to the KNN algorithm that computes the similarity distances between the coming sample with all others. As a result, it typically requires a much longer prediction time (see Table~\ref{tab:model-deployment}).


\begin{figure}[ht]
    \centering
    \vspace{-2mm}
    \vbox to 4.5cm{
    \includegraphics[width=.45\textwidth]{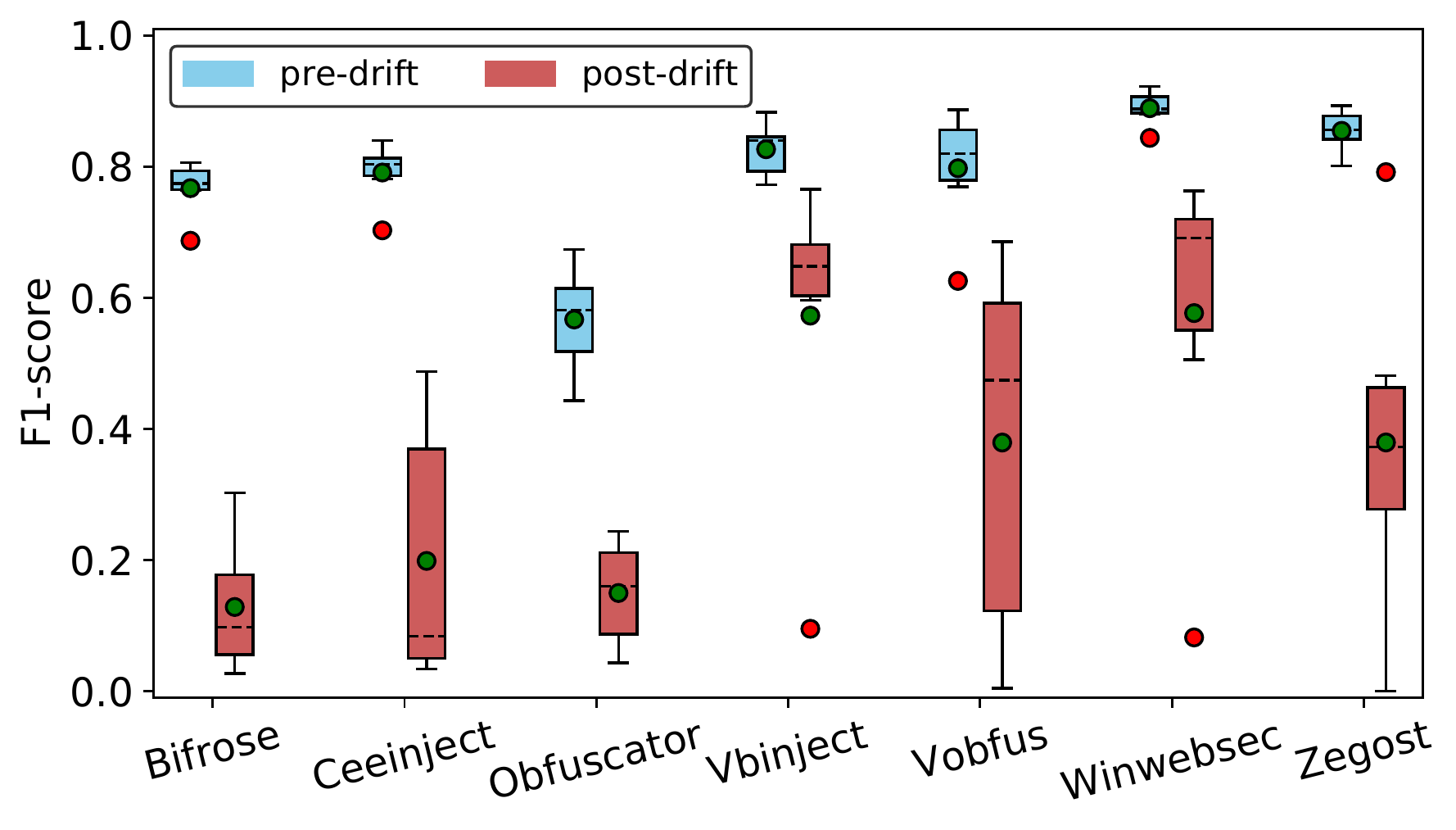}
    }
    \caption{The F1-score box-plot of the Pre-drift and the Post-drift samples for all methods on the MalwareDrift dataset.} 
    \label{drift-7-faimily}
\end{figure}

\vspace{-3mm}

\begin{figure}[htb]
\setlength{\abovecaptionskip}{0.cm}
\setlength{\belowcaptionskip}{-0.cm}
\vspace{-2mm}
    \begin{minipage}[t]{0.5\linewidth}
        \centering
        {\footnotesize  
         \vbox to 1.9cm{
         \includegraphics[width=0.8\textwidth]{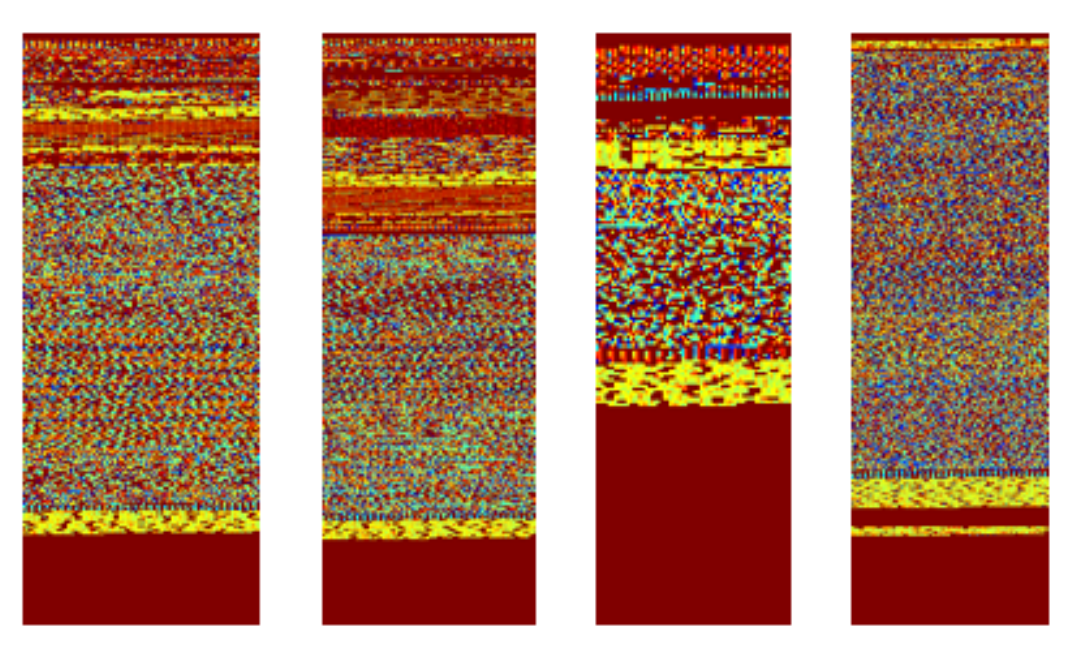}
        }
        }
        \subcaption{Vbinject pre-drift}
        \label{fig:Vbinject_pre}
    \end{minipage}%
    \begin{minipage}[t]{0.5\linewidth}
        \centering
        \vbox to 1.83cm{
        \includegraphics[width=0.9\textwidth]{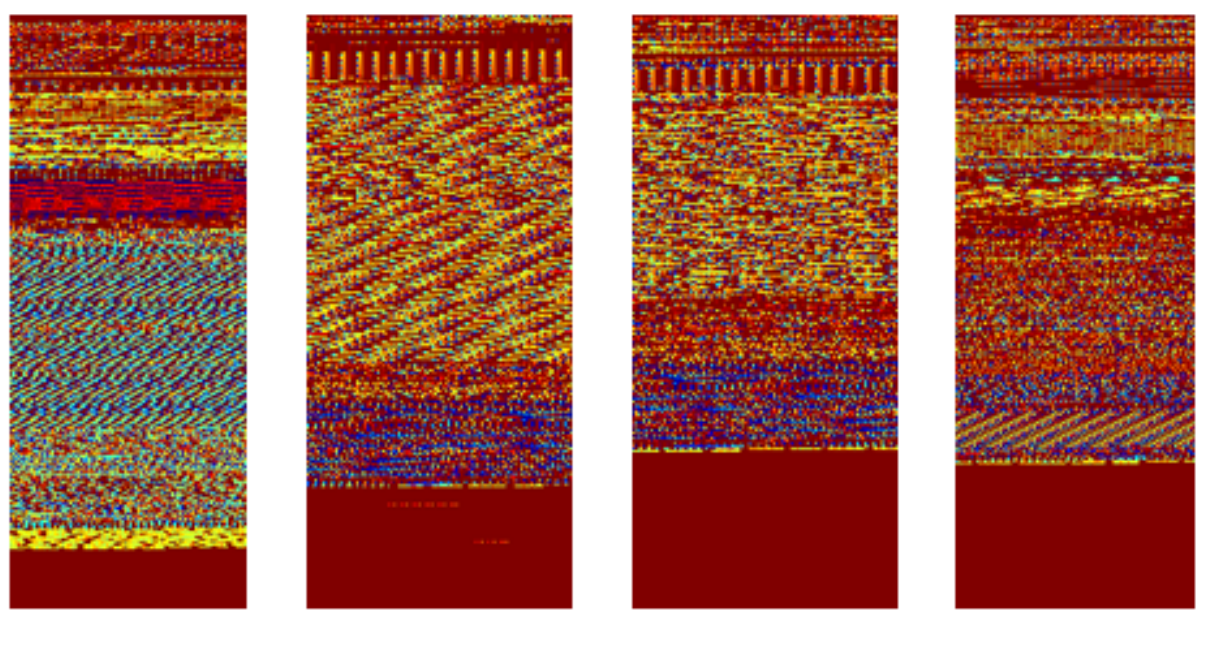}
        }
        \subcaption{Vbinject post-drift}
        \label{fig:Vbinject_post}
    \end{minipage}%
    
    \begin{minipage}[t]{0.5\linewidth}
        \centering
        \vbox to 1.1cm{
        \includegraphics[width=0.26\textwidth]{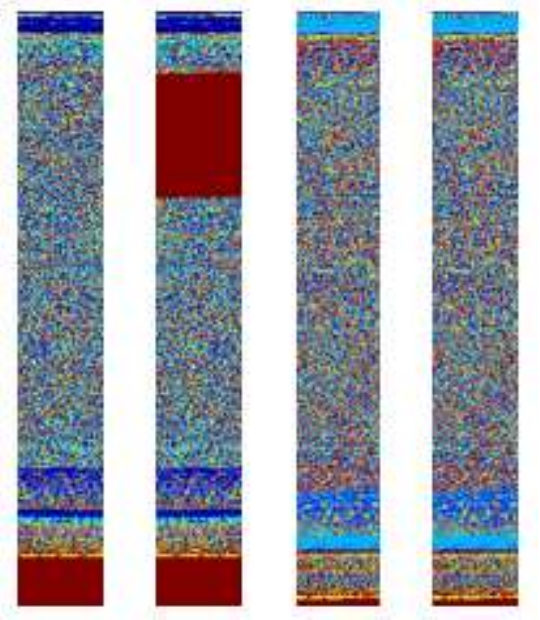}
        }
        \subcaption{Bifrose pre-drift}
        \label{fig:Bifrose_pre}
    \end{minipage}%
    \begin{minipage}[t]{0.5\linewidth}
        \centering
        \vbox to 1.1cm{
        \includegraphics[width=1.0\textwidth]{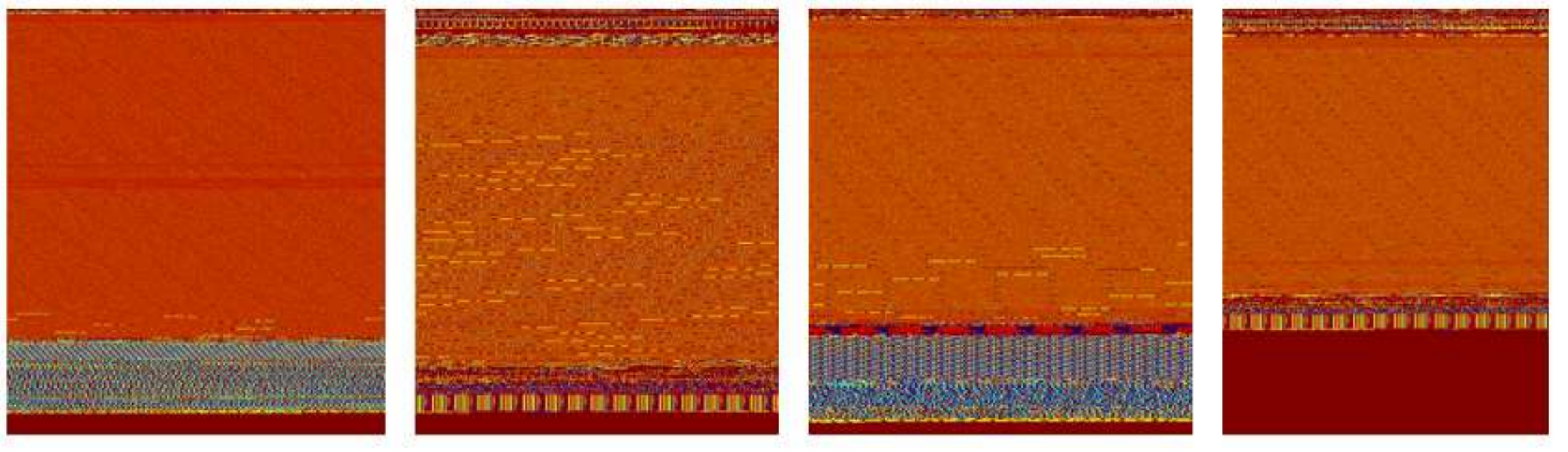}
        }
        \subcaption{Bifrose post-drift}
        \label{fig:Bifrose_post}
    \end{minipage}
\caption{The visualized images of the Pre-drift and Post-drift samples on the Vbinject and Bifrose family}
\label{fig:images}
\vspace{-4mm}
\end{figure}




\jjcmt{
Another interesting result is that CBOW+MLP, as one of the best in previous common machine learning scenarios, shows the sharpest decrease of nearly 70\%, and the classification performance on the post-drift dataset is only 10.88\%. 
CBOW+MLP extracts byte sequences from binary files, and focuses on the local context connection of byte sequences. However, unlike Opcode sequences, when concept drift occurs, the original byte sequence changes significantly with the changes of malware code, and thus previous local context connections almost no longer exist.}

\jjcmt{Results show that MCSC is the most robust to concept drift, with the least decrease by 27.07\% .
MCSC has two steps of feature extraction. After hashing the Opcode sequences to SimHash sequences, MCSC uses a convolution neural network to extract the key features of images converted from the SimHash sequences. 
Therefore, MCSC can potentially capture the fingerprints of similar families from the images based on Opcode sequences. 
Recall that CBOW+MLP also extracts Opcode sequence features, and IMCFN extracts image features based on original bytes, however, they are not robust to concept drift like MCSC.
Opcode sequences are robust to changes and it is more accurate to extract features from the image format. Therefore, the combination of the two steps shows the most robust performance against concept drift.}

\vspace{3mm}
\finding{CBOW+MLP shows the worst performance to concept drift, while MCSC is the most robust to concept drift. Results indicate that Opcode sequence preserves the family feature of PE malware and thus are effective and robust features for classification under concept drift.}

In order to investigate whether there is a significant difference on different PE malware families, we present the results in Figure~\ref{drift-7-faimily}. We can observe that all methods tend to perform consistently on pre-drift dataset, with an exception of Obfuscator due to code obfuscations, which disturbs the feature extraction and prediction performance of different methods. 
However, different methods show diverse performance on the post-drift dataset and this is especially obvious for the Ceeinject and Vobfus families. Moreover, all methods show consistent small performance drop on Vbinject. By a close look into the data, we find the major reason is that samples after concept drift in Vbinject do not show sharp changes compared to other families. For example, Figures~\ref{fig:Vbinject_pre} and~\ref{fig:Vbinject_post} show the visualized image of samples before and after the concept drift from Vbinject (named Group-1), while Figures~\ref{fig:Bifrose_pre} and~\ref{fig:Bifrose_post} show the samples from Bifrose (named Group-2), which obtains the largest performance drop. Comparing the images, we can observe that the impact of concept drift is relatively smaller in Group-1 than in Group-2. We further use the difference Hash similarity~\cite{dhash2017comparative} to measure the similarity between images before and after concept drift. The smaller the similarity value is, the more similar the images are. Finally, they are 26 and 32, respectively for Group-1 and Group-2, which indeed causes the big performance difference on different families.

\subsection{RQ3: Factors that Affect the Industry Usage of the Current Approaches}
\label{subsec:RQ3}
The ultimate goal of our work is to encourage deployment of the research models in real industry scenarios. Therefore, we conduct an interview with our industry partner, who is in charge of the virus detection product of a security company. In particular, we ask the following questions based on our study results. 
\begin{itemize}
    \item What classification methods are currently adopted in the company and why.
    \item What are the factors that affect choosing the suitable methods in real applicable scenario? 
    \item Is concept drift common in real application scenarios and what are the current status of handling concept drift? 
\end{itemize}

Currently, \textbf{two mainstream approaches are adopted in industry application scenario, i.e., the sandbox and the pattern based approaches.} Sandbox can be described as a virtual environment which execute the malware and extract runtime feature for classification. It is accurate yet is time and resource consuming. For instance, the sandbox can process around 5-10 samples per minute, which can be tolerated due to its high accuracy. Pattern based approaches are static detection methods which are based on the pattern/feature database. It is efficient in terms of time and resource consumption, yet is fragile to noises, obfuscation and concept drift.  

\textbf{Industry usage of the PE malware classification methods are mainly limited by three factors, i.e., the prediction precision and recall, the predicting time as well as the resource consumption, and the main resource concerns are runtime memory and CPU usage.} The first two factors decide the user experience and thus whether the corresponding method can be adopted, and the resource consumption decides what kind of devices can the methods be deployed on. As a concrete example, in one of their product which contains the learning-based malware classification model, they require the runtime memory to be below 1GB, which cannot be met by all of our studied methods. 
Another requirement is to be able to predict a malware within 0.1s with an accuracy above 93\%, which fillters out most of the binary-based and disassembly-based methods.  

\textbf{Concept drift usually happens due to malware evolving}, e.g., in scenarios where existing malware wants to escape detection, and there could be new non-kernel functionalities such as the communication and message passing techniques being changed. This happens frequently and raises challenges to malware family classification. There is a lack of specific mechanism to handle this case and current practice usually use the sandbox methods for such scenarios. 
Another observation is that in addition to concept drift, they also need to tackle with the challenge of the fast evolving new malware families and features. 
Except for the fine-grained family classification as defined by the existing academic datasets such as BIG-15, our industry partners are more interested in the detection of malware families based on their malicious behaviors, i.e., Trojan, Rootkit or Ransomware. However, there is a lack of research on this direction, highly likely due to the fact that there is no such datasets available. 

\finding{Real industry application scenario requires the classification methods to be able to tackle the challenges brought by the fast evolving of malware families. Moreover, there should be a trade-off between the resource consumption and prediction accuracy in consideration of the deployment environment and customer feedback. Therefore, the future research should focus more on (1) how to handle the fast evolving of malware family rather than only evaluate with one or a few dataset; (2) a more light-weight model with high prediction accuracy, and (3) malware family classification from the malicious behavior perspective.}

\section{Threats to Validity}
\label{sec:threats}


\textbf{Threats to internal validity} mainly lie in the implementation of different methods. In order to compare the results fairly, we employ the reported best configurations for each model if available, while for others we report the best results we have obtained after an intensive manual tuning process. We believe this strategy will mitigate the bias involved by different model settings.

Besides, due to the limit of physical memory, we set the batch size to 32 (originally 256) when training MalConv model, which may affect the results. However, we argue that it is reasonable and acceptable, especially in real application scenarios, where resources are critical according to the feedback from industry. In addition, we publish all our experimental data for replication and boosting future research.

\textbf{Threats to external validity} mainly lie in the selection bias of methods and datasets studied. In order to perform a systematic comparison, we have adopted 9 different methods, covering the mainstream image-based, binary-based and disassembly-based techniques. To alleviate the impact of datasets, we employ two commonly-used datasets (i.e., BIG-15 and Malimg), and  further construct two new datasets to reflect the latest progress of PE malware (MalwareBazaar) and the concept drift issue (MalwareDrift) from real industry practice.

\textbf{Threats to construction validity} mainly lie in the randomness and measurements in our experiment. To reduce the impact of randomness, we applied a 10-fold cross-validation in our comparing experiments on BIG-15, Malimg and MalwareBazaar datasets, and used 5-fold cross-validation in studying the effects of concept drift on various classification methods, rather than repeating each experiment several times. Macro-average is widely adopted to measure the performance of multi-class classification.

\section{Conclusion}
\label{sec:conclusion}

PE malware family classification has gained great attention and a large number of approaches have been proposed. 
In this paper, we first identify the gap of applying learning-based PE malware classification approaches in industry through a systematic empirical study, where we employ 9 different methods, covering the mainstream image-based, binary-based and disassembly-based techniques, and 4 different datasets for the experiment. Based on the obtained quantitative evaluation results in Section~\ref{subsec:RQ1},~\ref{subsec:RQ2}, and the requirements from industry (Section \ref{subsec:RQ3}), we conclude that: (1) There is no individual class of methods significantly outperforms the others; (2) All class of methods show performance degradation on concept drift, which is vital important in practice; (3) The prediction time and high memory consumption hinder existing approaches from being adopted for industry usage.
We further provide actionable guidance on future applied research: (1) focus more on how to handle the fast evolving of malware family; (2) explore light-weight models with high prediction accuracy; (3) take into account the malicious behavior features for malware family classification.

\section*{Acknowledgments}
This work is partially supported by 
the National Key Research and Development Program of China No. 2019QY1302; 
the NSFC-General Technology Basic Research Joint Funds under Grant U1836214; 
The NSFC Youth Funds under Grant 61802275;  
State Key Laboratory of Communication Content Cognition fund No. A32001.

\balance
\bibliographystyle{ACM-Reference-Format}
\bibliography{Reference}


\end{document}